\documentclass[twocolumn,showpacs,preprintnumbers,amsmath,amssymb]{revtex4}
\usepackage{graphicx}
\usepackage{textcomp}

\begin{document}
\draft
\title{Spectral singularities and Bragg scattering in complex crystals}
  \normalsize

\author{S. Longhi}
\address{Dipartimento di Fisica and Istituto di Fotonica e
Nanotecnologie del CNR, Politecnico di Milano, Piazza L. da Vinci 32, I-20133 Milano, Italy}


%
\bigskip
\begin{abstract}
\noindent  Spectral singularities that spoil the completeness of
Bloch-Floquet states may occur in non-Hermitian Hamiltonians with
complex periodic potentials. Here an equivalence is established
between spectral singularities in complex crystals and secularities
that arise in Bragg diffraction patterns. Signatures of spectral
singularities in a scattering process with wave packets are
elucidated for a $\mathcal{PT}$-symmetric complex crystal.
\end{abstract}



\maketitle

\section{Introduction}
The Dirac-von Neumann formulation of
quantum mechanics prescribes that the Hamiltonian $H$ of a physical
system must be Hermitian. This requirement ensures a real-valued
energy spectrum and a unitary (probability-preserving) temporal
evolution. In the last decade, complex extensions of quantum
mechanics which relax the Hermiticity constraint have been proposed
(see e.g. \cite{Bender07,Mostafazadehun}). Indeed, a diagonalizable
Non-Hermitian Hamiltonian (NHH) having a real spectrum is enough to
construct a unitary quantum system, provided that the inner product
of the Hilbert space is properly modified \cite{Mostafazadehun}. In
particular, Bender and collaborators showed that a NHH possessing
parity-time ($\mathcal{PT}$) symmetry may serve to develop a complex
extension of quantum mechanics below a phase transition
(symmetry-breaking) point \cite{Bender98,Bender02}. Experimental
realizations of $\mathcal{PT}$ Hamiltonians have been recently
proposed in optical media with a complex refractive index
\cite{Muga2,Makris08,Klaiman08,Shapiro09}, and the first observation
of $\mathcal{PT}$-symmetry breaking  has been reported
 in a passive optical waveguide coupler \cite{Guo09}.\\
For a NHH, the reality of the spectrum does not ensure
diagonalizability, which may be prevented by the presence of
exceptional points in the point-spectrum of $H$ \cite{ex1}, or of
spectral singularities in the continuous part of the spectrum
\cite{ss1,Samsonov05,Mostafazadeh09JPA}. Exceptional points, also
referred to as Hermitian degeneracies (see \cite{ex1}), correspond
to degeneracies where both eigenvalues and eigenvectors coalesce as
a system parameter is varied. Other singular points may also occur
in low-dimensional (e.g. matrix) non-Hermitian Hamiltonians, such as
branch points in the complex plane in which eigenvalue degeneracy do
not correspond to lack of completeness of the spectrum
\cite{Rotter05}. Such singularities have various physical
implications, which have been investigated in different physical
fields (see for instance \cite{Klaiman08,Guo09,ex1,ex2} and
references therein). They play an important role in the study of
open quantum systems, particularly in relation with the resonance
states (for a recent review see \cite{Rotter09}). But they must not
be confused with spectral singularities as defined in
\cite{ss1,Samsonov05,Mostafazadeh09JPA}. Unlike exceptional and
branch points that can be present for non-Hermitian operators with a
discrete spectrum, spectral singularities are exclusive features of
certain non-Hermitian operators having a continuous part in their
spectrum \cite{Mostafazadeh09JPA}. Following
\cite{ss1,Samsonov05,Mostafazadeh09JPA}, spectral singularities
refer to divergences (poles) of the resolvent on the continuous
spectrum of $H$, which do not correspond to square integrable
eigenfunctions. The physical implications of such spectral
singularities have been investigated so far for wave scattering from
complex potential barriers
\cite{Samsonov05,Mostafazadeh09JPA,Mostafazadeh09PRL,Mostafazadeh09PRA}
and shown -notably by Mostafazadeh- to correspond to resonance
states with vanishing spectral width
\cite{Mostafazadeh09PRL,Mostafazadeh09PRA}. Another physically
relevant class of NHHs with continuous spectrum is provided by
complex periodic potentials \cite{bande1,bande2,bande3}. Complex
crystals have been investigated in different areas of physics,
ranging from matter waves \cite{Keller97,Berry98,Berry98a} to optics
\cite{Berry98,Berry98a,Makris08,Shapiro09,Longhi09}. As compared to
ordinary crystals, complex crystals exhibit some unique properties,
such as violation of the Friedel's law of Bragg scattering, double
refraction, nonreciprocal diffraction, and anomalous transport
\cite{Makris08,Keller97,Berry98,Berry98a,Longhi09}, which make them
a rather unique class of synthetic materials. Spectral singularities
for periodic non-self-adjoint Schr\"{o}dinger operators have been
studied by mathematicians \cite{ss2}, however their
physical implications in connection to complex crystals have not been explored yet.\\
It is the aim of this work to investigate the onset of spectral
singularities in complex crystals and to show their physical impact
in Bragg scattering processes. In particular, we prove rather
generally that spectral singularities are associated to a secular
growth of plane waves diffracted off the crystal when the incident
angle is an integer multiple of the Bragg angle. With reference to a
specific $\mathcal{PT}$-symmetric complex crystal, previously
considered in Ref.\cite{Makris08}, we show that spectral
singularities occur at the $\mathcal{PT}$ symmetry breaking point
and can be revealed in a diffraction experiment using a
spatially-confined wave packet that excites the crystal at normal
incidence. Unlike plane wave excitation, broadening of the angular
spectrum for a wave packet is shown to lead to a saturation of the
secular growth of the wave amplitude.

\section{ Spectral singularities in complex crystals and Bragg scattering}
\subsection{Bloch-Floquet states and spectral singularities}
Wave dynamics in a complex one-dimensional crystal is governed by a
Schr\"{o}dinger-type equation, which in dimensionless form can be
written as
\begin{equation}
i \partial_t \psi(x,t)=-\partial^{2}_x \psi + V(x) \psi \equiv H
\psi
\end{equation}
where $V(x+a)=V(x)$ is the complex potential of period $a$.
Physically, Eq.(1) describes diffraction of matter or optical waves
by a complex lattice [see Fig.1(a)], where $t$ is a fictitious time
related to the propagation distance inside the crystal
\cite{Keller97,Berry98,Makris08,Longhi09}. The geometry of the
spectrum of $H$ is studied e.g. in \cite{bande3}; here we will
assume an entirely real energy spectrum, however at this stage no
requirement about $\mathcal{PT}$ symmetry of $H$ is needed. Like in
ordinary crystals, from the canonical form of the translation
operator \cite{AmJP} it follows that any eigenfunction $\phi(x)$ of
$H$ ($H \phi=E \phi$) which is bounded at $x \rightarrow \pm \infty$
is a linear combination of Bloch-Floquet type solutions satisfying
the condition $\phi(x+a)=\phi(x) \exp(iqa)$, where $ -\pi/a \leq q
<\pi/a$ is an arbitrary real number that varies in the first
Brillouin zone. The set of Bloch-Floquet eigenfunctions with energy
$E_{\alpha}(q)$ can be expanded in series of plane wave basis $|q,n
\rangle= (2 \pi)^{-1/2} \exp[i(q+nk_B)x]$ as
\begin{equation}
\phi_{\alpha}(x,q)=\sum_{n} w_n^{(\alpha)}(q) |q,n \rangle,
\end{equation}
where $k_B=2 \pi/a$ is the Bragg wave number, $\alpha$ is the band
index, and $n=0, \pm 1, \pm2,...$. In Eq.(2),
$\mathbf{w}^{(\alpha)}(q) \equiv \{ w_n^{(\alpha)}(q) \}$ are the
eigenvectors, with corresponding eigenvalues $E_{\alpha}(q)$, of the
matrix $\mathcal{H}(q)$ defined by
\begin{equation}
\mathcal{H}_{n,m}(q)=(q+nk_B)^2 \delta_{n,m}+V_{n-m}
\end{equation}
with $V(x)=\sum_{n} V_n \exp(ik_Bn)$. Let $\mathcal{J}(q)$ be the
Jordan canonical form of $\mathcal{H}(q)$, to which $\mathcal{H}(q)$
can be reduced by a similarity transformation, i.e.
$\mathcal{H}(q)=\mathcal{T}(q) \mathcal{J}(q) \mathcal{T}^{-1}(q)$.
In the band computation, degenerate eigenvalues of $\mathcal{H}(q)$
are counted by their geometric (and not algebraic) multiplicity,
i.e. by the number of Jordan blocks that represent the eigenvalue,
and the eigenvectors $\mathbf{w}^{(\alpha}(q)$ are thus linearly
independent. Note that the matrix representation of $H$ in the
plane-wave basis is given by $\langle q', n'|H q,n
\rangle=\mathcal{H}_{n',n}(q) \delta(q-q')$. Like in ordinary
crystals, the eigenvalues can be ordered such that
$E_{\alpha}(-q)=E_{\alpha}(q)$, and for any fixed value of $q$, with
$ q \neq 0, -\pi/a$, the energies $E_{\alpha}(q)$ are all distinct.
At $q=0$ or at $q= -\pi/a$, eigenvalue degeneracy, with algebraic
multiplicity not larger than 2, is allowed (see e.g. \cite{AmJP}).
For the problem of spectral singularities, a key role is played by
{\em defective} eigenvalues of $\mathcal{H}(q)$, at either $q=0$ or
$q= -\pi/a$, i.e. eigenvalues whose geometric multiplicity is
smaller than their algebraic multiplicity. More precisely, we will
prove that the completeness of a periodic NHH admitting a purely
continuous spectrum, i.e. the absence of spectral singularities in
complex crystals, is equivalent to the issue of completeness (i.e.
absence of defective eigenvalues) for the non-Hermitian matrix
Hamiltonian $\mathcal{H}(q)$ at $q=0$ and $q=-\pi/a$. Note that the
discrete NHH problem defined by the matrix $\mathcal{H}(q)$ at
$q=0,-\pi/a$ can be obtained from the original problem (1) provided
that the functional space is restricted to the space of functions
satisfying the periodic boundary conditions $\psi(x+a,t)=\pm
\psi(x,t)$. Moreover, the appearance of a defective eigenvalue of
$\mathcal{H}(q)$ (when a control parameter is varied) corresponds to
the occurrence of an exceptional point in the discrete spectrum of
$\mathcal{H}(q)$. The correspondence between spectral singularities
in the continuous spectrum of $H$ and exceptional points (defective
eigenvalues) in the point spectrum of the matrix $\mathcal{H}(q)$ is
more precisely established by
following theorem:\\
{\it Theorem 1.} The set of Bloch-Floquet eigenfunctions (2) is
complete, i.e. $H$ is diagonalizable, if and only if the matrix
$\mathcal{H}(q)$ does not have defective eigenvalues at $q=0$ or
$q=-\pi/a$. In other words, spectral singularities in the continuous
spectrum of $H$ correspond to exceptional points in the point
spectrum of the matrix $\mathcal{H}(q)$ at the center ($q=0$) or at
the edge ($q=-\pi/a$)
of the Brillouin zone.\\
{\it Proof.} Let us first prove that $\{ \phi_{\alpha}(x,q)\}$ is a
complete set of (improper) functions if and only if $\mathcal{H}(q)$
is diagonalizable for any arbitrary value of $q$ inside the first
Brillouin zone. Suppose that $\{ \phi_{\alpha}(x,q)\}$ is a complete
set. Then for any $q$ in the first Brillouin zone and for any
integer $n$, there exists a set of complex numbers $\{
c_{n,\alpha}(q) \}$ such that $|q,n \rangle=\sum_{\alpha}
c_{n,\alpha}(q) \phi_{\alpha}(x,q)$, i.e. $|q,n \rangle=\sum_{m}
\left( \sum_{\alpha} c_{n,\alpha}(q) w_{m}^{(\alpha)}(q) \right)
|q,m \rangle$, where we used Eq.(2). Hence $ \sum_{\alpha}
c_{n,\alpha}(q) w_{m}^{(\alpha)}(q) =\delta_{n,m}$. Since the
previous relation holds for any integer value $n$, the set of
eigenvectors $\{ \mathbf{w}^{(\alpha)}(q) \}$ is complete, i.e.
$\mathcal{H}(q)$ is diagonalizable. Suppose now that
$\mathcal{H}(q)$ is diagonalizable, and let us show that $\{
\phi_{\alpha}(q,x)\}$ is a complete set. In fact, for an arbitrarily
assigned square integrable function $f(x)$, let $F(k)$ be the
Fourier transform of $f(x)$ and $F_n(q)=F(q+nk_B)$, where $q$ varies
in the first Brillouin zone and $n=0, \pm 1, \pm 2,...$. Then
$f(x)=\sum_n \int_{-\pi/a}^{\pi/a} dq F_n(q) |q,n \rangle$. Since
$\mathcal{H}(q)$ is diagonalizable, the eigenvectors
$\mathbf{w}^{(\alpha)} (q)$ form a complete set and one can thus
determine a set of functions $c_{\alpha}(q)$ such that
$F_n(q)=\sum_{\alpha} c_{\alpha} (q) w_n^{(\alpha)}(q)$. Hence
$f(x)= \sum_{\alpha} \int_{-\pi/a}^{\pi/a} dq  c_{\alpha}(q) \sum_n
w_n^{(\alpha)}(q) |q,n \rangle$ $=\sum_{\alpha}
\int_{-\pi/a}^{\pi/a} dq c_{\alpha}(q) \phi_{\alpha}(x,q)$, i.e.
$f(x)$ can be decomposed as a superposition of Bloch-Floquet
eigenfunctions. Since  $f(x)$ is arbitrary, it follows that
$\{\phi_{\alpha}(x,q) \}$ is a complete set. The theorem is finally
proved after observing that, for $q \neq 0, -\pi/a$, the eigenvalues
$E_{\alpha}(q)$ are distinct, and thus $\mathcal{H}(q)$ is
diagonalizable.\par
 Similarly to spectral singularities found in
scattering complex potentials \cite{Mostafazadeh09JPA}, a physically
relevant consequence of spectral singularities is to prevent the
construction of a biorthogonal eigensystem for $H$, i.e. to resolve
the identity in terms of the biorthogonal basis associated to $H$.
By definition, $H$ is diagonalizable if $\phi_{\alpha}(x,q)$,
together with a set of (generalized) eigenfunctions
$\phi_{\alpha}^{\dag}(x,q)$ of the adjoint $H^{\dag}$, form a
complete biorthogonal system, i.e. they satisfy $\langle
\phi_{\alpha}(x,q) | \phi_{\beta}^{\dag}(x,q')
\rangle=\delta_{\alpha,\beta}\delta(q-q')$ and $\sum_{\alpha}
\int_{-\pi/a}^{\pi/a}dq | \phi_{\alpha}(x,q) \rangle  \langle
\phi_{\alpha}^{\dag}(x,q) | =\mathcal{I}$. As
$H^{\dag}=-\partial^2_x+V^*(x)$, it can be easily shown that one has
$\phi_{\alpha}^{\dag}(x,q)=\mathcal{N}_{\alpha}(q)
\phi_{\alpha}^{*}(x,-q)$, where $\mathcal{N}_{\alpha}(q)$ is a
multiplying term that needs to be determined. A direct computation
of the scalar product  $\langle \phi_{\alpha}(x,q)
|\phi_{\beta}^{\dag}(x,q') \rangle$ using expansion (2) yields
$\langle \phi_{\alpha}(x,q) |\phi_{\beta}^{\dag}(x,q') \rangle=
 \delta(q-q')
\delta_{\alpha,\beta} \mathcal{N}_{\alpha}(q)
\mathcal{D}_{\alpha}(q)$, where we have set
$\mathcal{D}_{\alpha}(q)=\langle \mathbf{w}^{(\alpha)}(q)|
\mathbf{w}^{(\alpha) \dag}(q) \rangle$ and $\mathbf{w}^{(\alpha)
\dag}(q)=\mathbf{w}^{(\alpha)*}(-q)$. Therefore,  completeness of
the biorthogonal system is ensured by letting
$\mathcal{N}_{\alpha}(q)=1/\mathcal{D}_{\alpha}(q)$  provided that
that $\mathcal{D}_{\alpha}(q) \neq 0$. Because
$\mathcal{D}_{\alpha}(q)$ vanishes if and only if the energy
$E_{\alpha}(q)$ is a defective eigenvalue of the matrix
$\mathcal{H}(q)$, from theorem 1 one concludes that a spectral
singularity prevents the
construction of a biorthogonal eigensystem for $H$.\\
Finally, it is also worth mentioning the following theorem, which
shows that the defective eigenvalues of $\mathcal{H}$ correspond to
the spectral singularities of $H$, defined as the divergence points
of
the resolvent $G(z)=(z-H)^{-1}$ on the continuous spectrum. \\
{\it Theorem 2}. Any defective eigenvalue $E$ of $\mathcal{H}$ is a
divergence point for the resolvent $G(z)=(z-H)^{-1}$.\\
{\it Proof.} Let $|\chi \rangle$ and $ | \varphi \rangle$ be two
 square integrable functions of the Hilbert space, and let us
 introduce the complex function of $z$, $G_{\chi,\varphi}(z)$,
 defined by $G_{\chi,\varphi}(z) \equiv \langle \chi | G(z) \varphi \rangle$. To
prove the theorem, it is enough to show that there exists at least a
couple of functions $|\chi \rangle$ and $ | \varphi \rangle$ such
that $G_{\chi,\varphi}(z)$ is unbounded as $z \rightarrow E_0$,
$E_0$ being a defective eigenvalue of $\mathcal{H}$. To this aim,
after expanding $|\chi \rangle$ and $|\varphi \rangle$ on the
plane-wave basis $|q,n \rangle$,  one can readily show that
\begin{equation}
G_{\chi,\varphi}(z)=\sum_{n,m} \int_{-\pi/a}^{\pi/a} dq
\chi^{*}_m(q) \varphi_n(q) \mathcal{R}_{m,n}(z,q),
\end{equation}
 where $\chi_m(q)
\equiv \langle q,m| \chi \rangle$, $\varphi_n(q) \equiv \langle q,n|
\varphi \rangle$ and
$\mathcal{R}(z,q)=[z-\mathcal{H}(q)]^{-1}=\mathcal{T}(q)[z-\mathcal{J}(q)]^{-1}
\mathcal{T}^{-1}(q)$ is the resolvent of the matrix
$\mathcal{H}(q)$. Note that the spectral functions $\chi_m(q)$ and
$\varphi_n(q)$ are simply related to the Fourier transforms of $|
\chi \rangle$ and $|\varphi \rangle$ as described in the proof of
Theorem 1, and hence the inversion relations $|\chi (x) \rangle =
\sum_n \int_{-\pi/a}^{\pi/a} dq \chi_n(q) |q,n \rangle$ and $|
\varphi (x) \rangle = \sum_n \int_{-\pi/a}^{\pi/a} dq \varphi_n(q)
|q,n \rangle$ hold. If $E_0$ is a defective eigenvalue of
$\mathcal{H}(q)$, say at $q=0$, with algebraic multiplicity 2 and
geometric multiplicity 1, then there exists one element of the
matrix $\mathcal{R}(z,q)$, say $\mathcal{R}_{m_0,n_0}(z,q)$, which
has a second-order pole at $z=E_0$ when $q=0$. This follows from the
well-known form of the resolvent $[z-\mathcal{J}(q)]^{-1}$ of a
Jordan matrix possessing a Jordan block with dimension 2. In the
neighborhoods of $q=0$ and $z=E_0$, $\mathcal{R}_{m_0,n_0}(z,q)$
behaves like $\sim 1/[(z-E_0-\alpha q)(z-E_0+\alpha q)]$, where
$\alpha$ is a constant and $E_0 \pm \alpha q$ are the two
eigenvalues that cross and become a defective eigenvalue at $q=0$.
Let us now choose the spectral functions $\chi_m(q)$ to vanish for
$m \neq m_0$,  and $\varphi_n(q)$ to vanish for $n \neq n_0$;
conversely, $\chi_{m_0}(q)$ and $\varphi_{n_0}(q)$ are assumed to be
nonvanishing and  bounded functions of $q$ in the interval $(-\pi/a,
\pi/a)$. The functions $| \chi \rangle$ and $| \varphi \rangle$ in
direct space that yield such spectral functions are determined by
the corresponding inversion relations given above. For instance, if
$\chi_{m_0}(q)$ and $\varphi_{n_0}(q)$ assume a constant value in
the interval $(-\pi/a, \pi/a)$, then $| \chi (x) \rangle=\mathcal{N}
[\sin(\pi x/a)/(x)] \exp(ik_Bm_0x)$ and $| \varphi (x)
\rangle=\mathcal{N} [\sin(\pi x/a)/(x)] \exp(ik_B n_0x)$, where
$\mathcal{N}$ is a normalization constant. Under such a choice, the
sum in Eq.(4) reduces to the only term $m=m_0$ and $n=n_0$. Owing to
the behavior of $\mathcal{R}(z,q)$ near $z=E_0$ and $q=0$, the
corresponding integral on the right-hand-side of Eq.(4) diverges as
$z \rightarrow E_0 \pm i0^+$, i.e. $G_{\chi,\varphi}(z)$ is
unbounded in the neighborood of $z= E_0$, which proves the theorem.
\\
\\
\subsection{Spectral singularities and secular Bragg diffraction} An
important physical implication of spectral singularities in complex
crystals is the appearance of a secular growth of the amplitudes of
waves scattered off the lattice when it is excited by a plane wave
at special incident angles. Such an anomalous behavior has been
previously noticed by Berry \cite{Berry98,Berry98a} for certain
absorptive potentials; here we prove that this is a very general
feature of complex crystals
related to the existence of spectral singularities. Namely:\\
{\it Theorem 3}. Let $\psi(x,t)=\exp(-iHt) \psi_0(x)$ be the Bragg
diffraction pattern corresponding to crystal excitation, at $t=0$,
with a plane wave $\psi(x,0)=\exp(ikx)$ of wave number $k$. Then $H$
has spectral singularities if and only if for some integer $n$ and
$k=nk_B/2$ the solution $\psi(x,t)$ contains secular (linearly
growing) terms in $t$.\\
{\it Proof.} Let us set $k=q+l_0 k_B$, where $l_0$ is an integer and
$ -\pi/a \leq q < \pi/a$. Then the solution $\psi(x,t)$ of Eq.(1)
with the initial condition $\psi(x,0)=\exp(ikx)$ is given by
$\psi(x,t)= \sum_{l=-\infty}^{\infty} c_l(t) \exp[i(q+lk_B)x]$,
where $c_l(t)$ satisfy the coupled equations $i(d c_l/dt)=\sum_m
\mathcal{H}_{l,m}(q)c_m(t)$ with the initial conditions
$c_l(0)=\delta_{l,l_0}$. Physically, the coefficients $c_l$ are the
amplitudes of diffracted waves at various orders at plane $t$ [see
Fig.1(a)]. Hence $c_l(t)=\mathcal{M}_{l,l_0}(q,t)$, where
$\mathcal{M}(q,t)= \exp[-i t \mathcal{H}(q)]$. The exponential
matrix $\mathcal{M}$ can be calculated from the Jordan decomposition
of $\mathcal{H}$ as $\mathcal{M}=\mathcal{T} \exp(-i t \mathcal{J})
\mathcal{T}^{-1}$. Having in mind the form of the exponential of a
Jordan matrix \cite{algebra}, it follows that secular growing terms
in some of the coefficients $c_l(t)$ [and hence in $\psi(x,t)$]
appear if and only if $\mathcal{H}(q)$ has at least one defective
eigenvalue, the largest growing term being $\sim t^{\rho}$ where
$\rho$ is the (maximal) difference between the algebraic and
geometric multiplicity of defective eigenvalues. In our case,
$\rho=1$ because the maximal algebraic multiplicity of any
eigenvalue of $\mathcal{H}(q)$ is 2; moreover, since
$\mathcal{H}(q)$ may have defective eigenvalues only for $q=0$ or
$q=-\pi/a$, secular growing terms may appear solely when the wave
number $k$ of the exciting plane wave is an integer multiple of
$k_B/2$. The absence of secular growing terms in $c_l(t)$ for any
excitation wave number $k=n k_B/2$ ($n=0, \pm 1, \pm 2,...$) implies
that the matrix $\mathcal{J}$ is diagonal, which ensures the lack of
defective eigenvalues of $\mathcal{H}$ and thus of
spectral singularities of $H$ according to Theorem 1.\\
\\
\section{Bragg scattering in $\mathcal{PT}$ complex crystals and wave packet dynamics}
Let us specialize the previous results to the case of a
$\mathcal{PT}$-symmetric lattice, for which $V(-x)=V^*(x)$. Let
$V_R(x)$ and $\lambda V_I(x)$ be the real and imaginary parts of the
potential, respectively, where $\lambda \geq 0$ measures the
anti-Hermitian strength of $H$. The spectrum of $H$ is real for
$\lambda \leq \lambda_c$, where $\lambda_c \geq 0$ defines the
symmetry breaking point. According to the previous analysis,
complex-conjugate pairs of eigenvalues for $\mathcal{H}(q)$ should
appear as $\lambda$ is increased from below to above $\lambda_c$.
The typical scenario that describes symmetry breaking in a
finite-dimensional $\mathcal{PT}$ matrix is the appearance of an
exceptional point via the merging of two real eigenvalues into a
single real and defective eigenvalue at $\lambda=\lambda_c$ (see,
for instance, \cite{Klaiman08,Weigert}). We note that the existence
of such a branching point for a certain class of non-hermitian
matrices  as a control parameter is varied was proven in a rather
generally way in Ref.\cite{Moyseiev80} (see also \cite{Klaiman08}).
We may thus conjecture that for a $\mathcal{PT}$ complex crystal
symmetry-breaking is accompanied by the appearance of spectral
singularity, which arise from defective eigenvalues of $\mathcal{H}$
at $q=0$ or $q=-\pi/a$. This scenario is in agreement with numerical
or analytical results obtained from band computation of specific
complex periodic potentials (see, for instance,
\cite{Makris08,Shapiro09}).\\
\begin{figure}[htbp]
  \includegraphics[width=85mm]{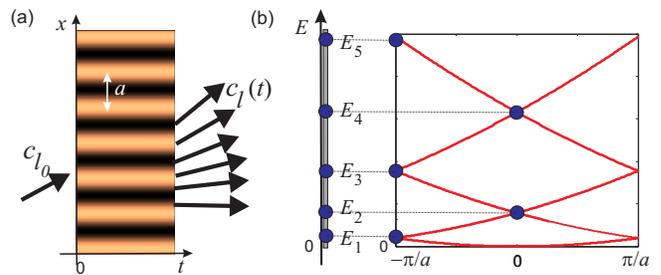}\\
   \caption{(color online) (a) Schematic of Bragg scattering of a plane wave off a complex crystal. (b) Band structure of the
the complex crystal $V(x)=V_0 \exp(ik_Bx)$. The circles mark the
spectral singularities inside the continuous spectrum.}
\end{figure}
As an example, let us consider the $\mathcal{PT}$ crystal defined by
\begin{equation}
V_R(x)=V_0 \cos(2 \pi x /a) \; , \; \;  V_I(x)=V_0 \sin(2 \pi x /a),
\end{equation}
 which has been recently considered to highlight
unusual diffraction and transport properties of complex optical
lattices \cite{Makris08,Longhi09}. In this case, $\lambda_c=1$
\cite{Makris08} and at the symmetry breaking point one has $V(x)=V_0
\exp(ik_Bx)$, a potential which is amenable for an analytical study
\cite{Berry98,bande1}. For this potential, $\mathcal{H}$ has a block
diagonal form, namely $\mathcal{H}_{n,m}=(n+k_Bq)^2 \delta_{n,m}+V_0
\delta_{m,n+1}$, and its eigenvalues are simply the elements on the
main diagonal, i.e. $E_{\alpha}(q)=(q+\alpha k_B)^2$ ($\alpha=0, \pm
1, \pm 2,...)$. This means that, as previously noticed
\cite{bande1,Makris08,Longhi09}, the band structure of the
$\mathcal{PT}$ potential $V=V_0 \exp(ik_B x)$ coincides with the
free-particle energy dispersion curve $E=k^2$, periodically folded
inside the first Brillouin zone [see Fig.1(b)]. The eigenvalues are
distinct for $q \neq 0, -\pi/a$. At the crossings of the folded
parabolas of Fig.1(b), i.e. at $q=0$ and at $q= -\pi/a$, one has
$E_{-\alpha}(q)=E_{\alpha}(q)$ and $E_{1-\alpha}(q)=E_{\alpha}(q)$,
respectively, i.e. the eigenvalues coalesce in pairs and become
defective. Therefore, the continuous spectrum of $H$, $E \geq 0$,
contains a sequence of spectral singularities at $E_n=( n k_B/2)^2$,
$n =1,2,3,...$ [see Fig.1(b)], which spoil the completeness of the
Bloch-Floquet eigenfunctions. The defective nature of degenerate
eigenvalues at $q=0$ and $q= -\pi/a$ can be readily proven by direct
calculation of the eigenvectors $\mathbf{w}^{(\alpha)}$ of
$\mathcal{H}$. Note that, as wave scattering from complex potential
{\it barriers} enables a finite number of spectral singularities in
the continuous spectrum \cite{Samsonov05,Mostafazadeh09PRL}, in our
example the number of spectral singularities is countable but
infinite.\\
According to Theorem 3, a secular growth of Bragg diffraction
pattern for a plane wave that excites the crystal at normal
incidence (or tilted by an angle which is an integer multiple of the
Bragg angle) provides a distinctive signature of the appearance of
spectral singularities at the $\mathcal{PT}$ symmetry-breaking
transition point $\lambda=\lambda_c=1$. However, in any experimental
setting aimed to observe such a secular growth, the wave that
excites the crystal is always spatially limited or truncated, and it
is thus of major relevance to investigate the impact of spectral
singularities on the evolution of a wave packet with a broadened
angular spectrum, an issue which was not considered in previous
works by Berry. Here we investigate the Bragg diffraction of a wave
packet with a broadened momentum distribution, $\psi(x,0)=\int dk
F(k) \exp(ikx)$, and show that spectral broadening leads to a {\it
saturation} of the secular growth of scattered waves. Such a
saturation behavior is basically due to the fact that spectral
singularities are of measure zero (they are a countable set of
points embedded in the continuous energy spectrum $E \geq 0$). For
the sake of simplicity, we consider a shallow lattice and a wave
packet with a narrow spectrum $F(k)$ centered at $ k=-k_B/2$ of
width $\Delta k \ll k_B$. Following the same lines detailed in the
proof of Theorem 3, one can show that the diffraction pattern
$\psi(x,t)$ can be written as the interference of different wave
packets describing one-side diffraction at various orders, namely
\begin{equation}
\psi(x,t)=\psi_0(x,t)+V_0 \psi_1(x,t)+V_0^2 \psi_2(x,t)+...
\end{equation}
where $\psi_n(x,t)= \int dk F(k) c_n(k,t) \exp[i(k+nk_B)x]$ with
$c_0(k,t)=\exp(-ik^2 t)$ and $c_n(k,t)=-i \int_0^t d \xi
c_{n-1}(k,\xi) \exp[i(k+nk_B)^2(\xi-t)]$ for $n \geq 1$. For a
shallow lattice, i.e. for $|V_0| \ll 1$, we can limit to consider
the first two terms on the right hand side of Eq.(6). The leading
term, $\psi_0(x,t)=\int dk F(k) \exp(-ik^2 t+ikx)$, is simply the
freely-diffracting wave packet that one would observe in the absence
of the crystal and that propagates with a constant speed
$v=dx/dt=k_B$. The expression of $\psi_1(x,t)$ is more involved,
however its asymptotic behavior for $t \gg \pi /(k_B  \Delta k)$
assumes a rather simple and physically interesting form, namely
\begin{equation}
\psi_1 \sim \ -\frac{i \pi}{k_B} F\left( -\frac{k_B}{2} \right) \exp
\left(i \frac{k_B x}{2}-i \frac{k_B^2 t}{4} \right) \Phi \left(
\frac{x}{k_B t} \right)
\end{equation}
\begin{figure}[htbp]
  \includegraphics[width=85mm]{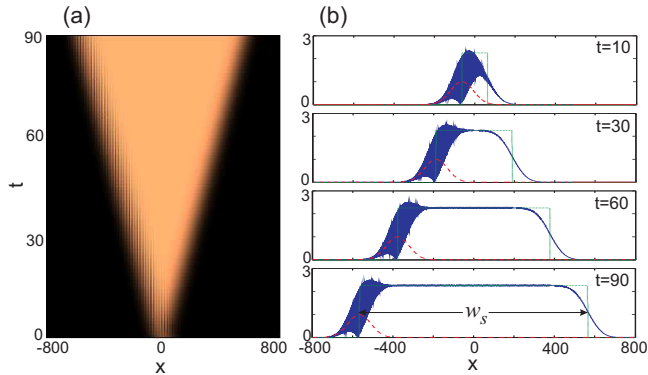}\\
   \caption{(color online) Evolution of a Gaussian wave packet in the complex lattice $V(x)=V_0 \exp(ik_Bx)$ for $V_0=0.2$,
   $a=1$ and $w=80$. In (a) a snapshot of $|\psi(x,t)|$ is shown,
   whereas in (b) the profiles of $|\psi(x,t)|$ at a few values of
   $t$ are reported. In (b), the dashed lines correspond to the wave
   packet evolution in absence of the lattice, whereas the dotted
   curves correspond to $|\psi_1(x,t)|$ as predicted by Eq.(7).}
\end{figure}
\begin{figure}[htbp]
  \includegraphics[width=80mm]{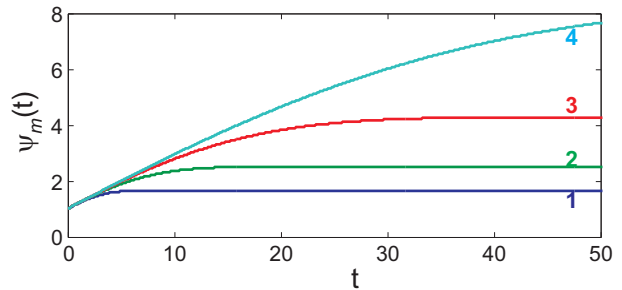}\\
   \caption{(color online) Saturation of secular Bragg scattering for a Gaussian wave packet that excites the complex lattice of Fig.2 at normal
   incidence. The curves  show the evolution of the maximum wave packet amplitude $\psi_m(t)$ versus $t$ for a few values
   of input wave packet spot size $w$. Curve 1: $w=40$; curve 2: $w=80$; curve 3:
   $w=150$; curve 4: $w=300$.}
\end{figure}
where $\Phi(\xi)=1$ for $|\xi|<1$ and $\Phi(\xi)=0$ for $|\xi|>1$.
Equation (7) shows that, owing to wave packet broadening in momentum
space, the secular growth with $t$ of the diffracted beam saturates
to the value $\sim (\pi/k_B)|F(-k_B/2)|$ while the beam assumes a
square shape whose width $w_s$ spreads in space with a constant
speed $v=dw_s/dt=k_B$. Note that $v$ equals the translation speed of
the freely-diffracting wave packet $\psi_0$. This behavior is
confirmed by direct numerical simulations of Eq.(1), as shown in
Fig.2. The figure depicts the evolution of $|\psi(x,t)|$ for a
Gaussian wave packet $\psi(x,0)=\exp[-(x/w)^2-ik_Bx/2]$ with
spectrum $F(k)=[w /(2 \sqrt \pi)] \exp[-(k+k_B/2)^2 w^2/4]$ that
excites the crystal at $t=0$. The evolution of the freely
diffracting wave packet $|\psi_0|$ and of the asymptotic behavior of
$|\psi_1|$ predicted by Eq.(6) are also shown for comparison. Note
the formation of interference fringes on the left side of the wave
packet, which arise from the interference of $\psi_0$ and $\psi_1$.
The spectral-broadening-induced saturation of the secular growth of
the wave packet is clearly shown in Fig.3, where the behavior of the
maximum amplitude $\psi_m(t)$ of the wave packet versus $t$, defined
by
\begin{equation}
 \psi_m(t) = \mathrm{max}_x |\psi(x,t)|,
\end{equation}
\begin{figure}[htbp]
  \includegraphics[width=85mm]{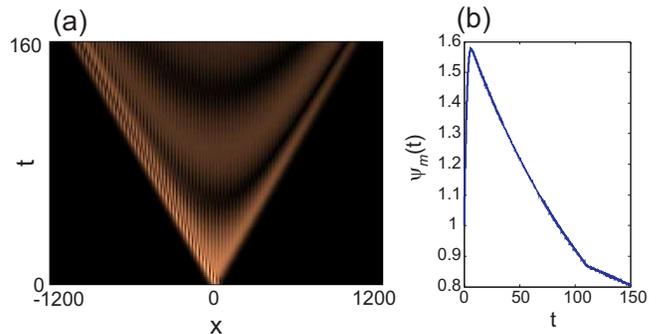}\\
   \caption{(color online) (a) Evolution of a Gaussian wave packet [modulus of $\psi(x,t)$] in the complex lattice
   (5) for $V_0=0.2$, $a=1$, $w=80$, and below the $\mathcal{PT}$ symmetry breaking point ($\lambda=0.9$). In (b), the
   corresponding evolution of the maximum wave packet amplitude $\psi_m(t)$ is shown.}
\end{figure}
\begin{figure}[htbp]
  \includegraphics[width=85mm]{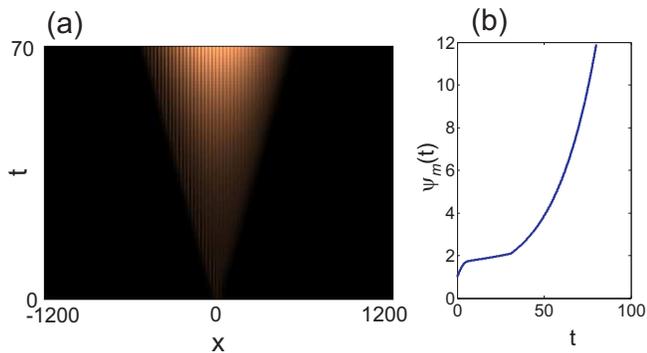}\\
   \caption{(color online) Same as Fig.4, but above the $\mathcal{PT}$ symmetry-breaking point ($\lambda=1.1$).}
\end{figure}
 is depicted for a
few decreasing values of the wave packet input spot size $w$. Note
that in the early stage of the dynamics the peak amplitude linearly
increases with $t$, as expected for a plane wave according to
Theorem 3. The linear growth then saturates to a steady state value
and, as Fig.3 clearly shows, the saturation process occurs earlier
for wave packets with smaller input size $w$. It should be pointed
out that saturation of $\psi_m(t)$ to a steady-state value, as
predicted by Eq.(7) and confirmed by numerical simulations depicted
in Figs.2 and 3, is indeed a signature of spectral singularities of
the underlying Hamiltonian that arise at $\lambda=\lambda_c=1$. This
is clearly shown in Figs.4 and 5, where a typical wave packet
evolution is reported for the complex crystal defined by Eq.(5)
either below ($\lambda=0.9 \lambda_c=0.9$, Fig.4) and above
($\lambda=1.1 \lambda_c=1.1$, Fig.5) the $\mathcal{PT}$ symmetry
breaking transition point. In both cases, $H$ is diagonalizable and
there are not spectral singularities. Note that below the
$\mathcal{PT}$ symmetry breaking point, the energy spectrum is
real-valued and, after an initial increase, $\psi_m(t)$ does not
settle down to a steady-state value, rather it tends to
monotonically decay (Fig.4) like in an ordinary (Hermitian) crystal.
On the other hand, above the $\mathcal{PT}$
 symmetry-breaking point ($\lambda=1.1$), the energy spectrum contains pairs of
complex-conjugate eigenvalues, and $\psi_m(t)$ monotonically
increases, as shown in Fig.5. Therefore, the scattering process with
wave packets can be used to highlight the appearance of spectral
singularities at the $\mathcal{PT}$ symmetric breaking point, which
are revealed as the saturation of the wave packet amplitude growth
to a steady-state (non-decaying) value. If the lattice is realized
in a periodic dielectric medium as discussed in \cite{Makris08}, at
light wavelength $\lambda=1.5 \; \mu$m, assuming a bulk refractive
index $n_s=1.5$ and a maximum refractive index change (both real and
imaginary parts) of $\Delta n=2 \times 10^{-4}$, the simulations of
Fig.2 correspond, as an example, to an optical lattice with spatial
period $a \simeq 6.2 \mu$m excited by a Gaussian beam of size $w
\simeq 493 \; \mu$m; the spatial units along the $x$ and $t$ axes
are $l_x \simeq 6.2 \; \mu$m and $l_t \simeq 477 \; \mu$m,
respectively. The same scales hold for the simulations shown in
Figs.4 and 5.
\\
\\
\section{Conclusions} In this work, it has been shown
that Bragg scattering in complex crystals provides a physically
important process to visualize the appearance of spectral
singularities in non-Hermitian Hamiltonians with complex periodic
potentials. In particular, clear signatures of spectral
singularities could be gained in a diffraction experiment with wave
packets. These results may suggest the investigation of possible
experimental systems to observe spectral singularities in complex
lattices, such as photonic systems \cite{Makris08,Guo09}.


\begin{thebibliography}{31}

\bibitem{Bender07}
C.M. Bender, Rep. Prog. Phys. {\bf 70}, 947 (2007).

\bibitem{Mostafazadehun}
A. Mostafazadeh, arXiv:0810.5643.


\bibitem{Bender98}
C. M. Bender and S. Boettcher, Phys. Rev. Lett. {\bf 80}, 5243
(1998).

\bibitem{Bender02}
C. M. Bender, D.C. Brody, and H. F. Jones, Phys. Rev. Lett. {\bf
89}, 270401 (2002); {\bf 92}, 119902(E) (2004).

\bibitem{Muga2}
A. Ruschhaupt, F. Delgado, and J.G. Muga, J. Phys. A {\bf 38}, L171
(2005).

\bibitem{Makris08}
K.G. Makris, R. El-Ganainy, D.N. Christodoulides, and Z.H.
Musslimani, Phys. Rev. Lett. {\bf 100}, 103904 (2008).

\bibitem{Klaiman08}
S. Klaiman, U. G\"{u}nther, and N. Moiseyev, Phys. Rev. Lett. {\bf
101}, 080402 (2008).

\bibitem{Shapiro09}
O. Bendix, R. Fleischmann, T. Kottos, and B. Shapiro, Phys. Rev.
Lett. {\bf 103}, 030402 (2009).

\bibitem{Guo09}
A. Guo, G.J. Salamo, D. Duchesne, R. Morandotti, M. Volatier-Ravat,
V. Aimez, G. A. Siviloglou, and D. N. Christodoulides, Phys. Rev.
Lett. {\bf 103}, 093902 (2009).

\bibitem{ex1}
 M.V. Berry, Czech. J. Phys. {\bf 54}, 1039 (2004); W.D. Heiss, J. Phys. A {\bf 37}, 2455
 (2004).

\bibitem{ss1}
M. A. Naimark, Amer. Math. Soc. Transl. {\bf 16}, 103 (1960); V. E.
Lyance, Amer. Math. Soc. Transl. {\bf 60}, 185 (1967).

\bibitem{Samsonov05}
B. F. Samsonov, J. Phys. A {\bf 38}, L397 (2005).

\bibitem{Mostafazadeh09JPA}
A. Mostafazadeh and H. Mehri-Dehnavi, J. Phys. A {\bf 42}, 125303
(2009).

\bibitem{Rotter05}
I. Rotter and A.F. Sadreev, Phys. Rev. E {\bf 71}, 036227 (2005); I.
Rotter, Czech. J. Phys. {\bf 55}, 1167 (2005).

\bibitem{ex2}
E. Persson, I. Rotter, H.-J. St\"{o}ckmann, and M. Barth, Phys. Rev.
Lett. {\bf 85}, 2478 (2000); P. Cejnar, S. Heinze, and M. Macek,
Phys. Rev. Lett. {\bf 99}, 100601 (2007); H. Cartarius, J. Main, and
G. Wunner, Phys. Rev. Lett. {\bf 99}, 173003 (2007).

\bibitem{Rotter09}
I. Rotter, J. Phys. A {\bf 55}, 1167 (2005).

\bibitem{Mostafazadeh09PRL}
A. Mostafazadeh, Phys. Rev. Lett. {\bf 102}, 220402 (2009).

\bibitem{Mostafazadeh09PRA}
A. Mostafazadeh, Phys. Rev. A {\bf 80}, 032711 (2009).

\bibitem{Keller97}
C. Keller, M.K. Oberthaler, R. Abfalterer, S. Bernet, J.
Schmiedmayer, and A. Zeilinger, Phys. Rev. Lett. {\bf 79}, 3327
(1997).

\bibitem{Berry98}
M.V. Berry, J. Phys. A {\bf 31}, 3493 (1998).

\bibitem{Berry98a}
M.V. Berry and D.H.J. O'Dell, J. Phys. A {\bf 31}, 2093 (1998).

\bibitem{bande1}
C.M. Bender, G.V. Dunne, P.N. Meisinger, Phys. Lett. A {\bf 252},
272 (1999).

\bibitem{bande2}
 H.F. Jones, Phys. Lett. A {\bf 262}, 242 (1999); Z.
Ahmed, Phys. Lett. A {\bf 286}, 231 (2001); J.M. Cervero, Phys.
Lett. A {\bf 317}, 26 (2003); A. Khare and U. Sukhatme, J. Math.
Phys. {\bf 47}, 062103 (2006); T. Curtright and L. Mezincescu, J.
Math. Phys. {\bf 48}, 092106 (2007).

\bibitem{bande3}
L.A. Pastur and V.A. Tkachenko, Math. Notes {\bf 50}, 1045 (1991);
 K.C. Shin, J. Phys. A {\bf 37}, 8287 (2004).

\bibitem{Longhi09}
S. Longhi, Phys. Rev. Lett. {\bf 103}, 123601 (2009).

\bibitem{AmJP}
A.A. Cottey, Am. J. Phys. {\bf 39}, 1235 (1971).

\bibitem{ss2}
See, for instance, O.A. Veliev, Russian J. Math. Phys. {\bf 13}, 101
(2006) and references therein.

\bibitem{algebra}
S. K. Godunov, {\it Ordinary differential equations with constant
coefficients} (American Mathematical Society, Providence, R.I,
1997), pp.43-49.

\bibitem{Weigert}
S. Weigert, J. Phys. A {\bf 39}, 235 (2006).

\bibitem{Moyseiev80}
N. Moiseyev and S. Friedland, Phys. Rev. A {\bf 22}, 618 (1980).


\end{thebibliography}

\end{document}